# Bayesian Source Separation and Localization

Kevin H. Knuth[a, b]

[a] Dept. of Neuroscience, Albert Einstein College of Medicine, Bronx NY
[b] Dept. of Speech and Hearing Sciences, CUNY Graduate Center, NY NY

**ABSTRACT**

The problem of mixed signals occurs in many different contexts; one of the most familiar being acoustics. The forward problem in acoustics consists of finding the sound pressure levels at various detectors resulting from sound signals emanating from the active acoustic sources. The inverse problem consists of using the sound recorded by the detectors to separate the signals and recover the original source waveforms. In general, the inverse problem is unsolvable without additional information.

This general problem is called source separation, and several techniques have been developed that utilize maximum entropy, minimum mutual information, and maximum likelihood. In previous work, it has been demonstrated that these techniques can be recast in a Bayesian framework. This paper demonstrates the power of the Bayesian approach, which provides a natural means for incorporating prior information into a source model. An algorithm is developed that utilizes information regarding both the statistics of the amplitudes of the signals emitted by the sources and the relative locations of the detectors. Using this prior information, the algorithm finds the most probable source behavior and configuration. Thus, the inverse problem can be solved by simultaneously performing source separation and localization.

It should be noted that this algorithm is not designed to account for delay times that are often important in acoustic source separation. However, a possible application of this algorithm is in the separation of electrophysiological signals obtained using electroencephalography (EEG) and magnetoencephalography (MEG).

## 1. INTRODUCTION

Perhaps the most familiar example of source separation occurs in the context of a listener attempting to pay attention to one speaker in an environment filled with many other sound sources. This problem is known as the Cocktail Party Problem [1] and as we well know the brain typically does an excellent job in focusing on one particular sound under a variety of conditions. Another important source separation problem appears in the context of an experimenter recording electromagnetic signals emitted by a particular neural source in a human brain while many other brain processes are producing additional signals. Typically in these problems, there is not sufficient information to deduce the source behavior. Instead, one is required to rely on a procedure of inference. This inference can depend on whatever additional information is available. To a person listening to a speaker, visual information obtained about the speaker's mouth movements is sometimes very helpful. Artificial sound separation algorithms typically rely only on prior information regarding the statistics of the amplitudes of particular sound signals. In contrast, neuroelectromagnetic source localization techniques typically only utilize prior information about the possible locations of the sources with respect to the detectors and the nature of the propagation of the signals from the sources to the detectors.

In this paper, we will demonstrate that these inference problems can be handled quite generally using a Bayesian formulation, and many types of prior information can be included to aid in the solution of a problem. In section 2, we will discuss how the Bayesian formalism can be used to treat these problems in general. A specific example, the Bell-Sejnowski Independent Component Analysis (ICA) algorithm,[2] which uses only prior information on signal statistics, will be derived in section 3. In section 4, we demonstrate how the Bayesian formulation allows one to incorporate specific prior information about the source geometry and mixing properties. The Bayesian separation and localization (BSL) algorithm is demonstrated in section 5 and compared to the Bell-Sejnowski ICA algorithm. Finally, in section 6, we demonstrate how one can obtain the desired source model parameters after the signals have been separated.

Further Author Information:
Email: knuth@email.arc.nasa.gov

## 2. APPLICATION OF BAYESIAN TECHNIQUES TO THE SEPARATION PROBLEM

The typical separation problem consists of the following elements: a set of sources emitting signals of some form, a medium through which the signals travel, and a set of detectors in this medium which record mixtures of the signals. It is generally assumed that there are *n* independent sources emitting signals $s_1(t), s_2(t), ..., s_n(t)$ and one observes an equal number of independent mixtures $x_1(t), x_2(t), ..., x_n(t)$, where the mixing is assumed to be linear and instantaneous. This linear mixing operation can be written in a compact form,

$$\mathbf{x}(t) = \mathbf{A}\,\mathbf{s}(t). \tag{1}$$

It should be noted that the assumptions made regarding instantaneous mixing and the fact that there are an equal number of sources and detectors, which makes $\mathbf{A}$ square, are not necessary for applying the Bayesian formalism, but instead greatly simplify the mathematics.

The difficulty of the problem is related to the fact that neither the matrix $\mathbf{A}$, the mixing matrix, nor the source signals is known. The problem is inherently an inference problem and is unsolvable without the inclusion of some additional prior information. To solve this problem, one must first choose a source model, which describes the variables of interest. The model could be simply one or both of the unknowns in Equation (1) above, or additional parameters on which the mixing matrix $\mathbf{A}$, or the source signals $\mathbf{s}(t)$ depend. Bayes' Theorem allows one to write the probability that a particular source model is correct in terms of the likelihood of the data and additional prior probabilities:

$$P(\mathbf{A},\mathbf{s}\,|\,\mathbf{x},I) \propto P(\mathbf{x}\,|\,\mathbf{A},\mathbf{s},I)\,P(\mathbf{A}\,|\,\mathbf{s},I)\,P(\mathbf{s}\,|\,I), \tag{2}$$

where *I* represents any prior information about the problem. Typically, the properties of the propagation of the signals through the medium do not depend on the source signals or their magnitudes, so we can simplify the first prior probability:

$$P(\mathbf{A},\mathbf{s}\,|\,\mathbf{x},I) \propto P(\mathbf{x}\,|\,\mathbf{A},\mathbf{s},I)\,P(\mathbf{A}\,|\,I)\,P(\mathbf{s}\,|\,I). \tag{3}$$

The probability of the model, or the degree to which we believe it is correct, is thus proportional to the product of the likelihood of the data given the model, and a product of the prior probabilities of the mixing matrix and the source signals based on any prior information. One can view Equation (3) as describing how the acquisition of some new information changes what we believe about the model.

Since there is a relationship between the data, the mixing matrix, and the source signals, given by Equation (1), it is not necessary to solve for both $\mathbf{A}$ and $\mathbf{s}(t)$. In fact, since the dimensions of $\mathbf{A}$ are much smaller, we can find the probability of $\mathbf{A}$ alone by treating $\mathbf{s}(t)$ as a nuisance parameter and by marginalizing over $\mathbf{s}(t)$, giving

$$P(\mathbf{A}\,|\,\mathbf{x},I) \propto P(\mathbf{A}\,|\,I)\,\int d\mathbf{s}\,P(\mathbf{x}\,|\,\mathbf{A},\mathbf{s},I)\,P(\mathbf{s}\,|\,I). \tag{4}$$

At this point, we have made no assumptions about the specific problem, nor have we included any prior information. The likelihood term is easily handled since we have assumed that the mixtures are linear and instantaneous. The first prior term describes our prior knowledge about the form of the mixing matrix. This knowledge includes information about the propagation of the signals through the medium, also known as the transfer function or the forward problem, and any information about the geometry of the sources and detectors. This is typically the type of information used by electromagnetic source location techniques.[3] The second prior term describes our knowledge about the form of the source signals. This prior is typically the focus of the blind source separation techniques.[2, 4, 5, 6]

## 3. BLIND SOURCE SEPARATION

We now will briefly outline the derivation of the Bell-Sejnowski ICA algorithm from the probability of the model described in Equation (4). Similar derivations can be found elsewhere.[7, 8, 9] We begin with Equation (4) and make the following assumptions. First, we assume that the problem is that of blind source separation, where we know nothing about the mixing process and have only minimal knowledge regarding the source signals. This lack of knowledge about the mixing process is reflected in our ignorance on the form of the mixing matrix $\mathbf{A}$, and values of the particular matrix elements $A_{ij}$. Assigning a uniform prior to the probability $P(\mathbf{A}\,|\,I)$, expresses this ignorance:

$$P(\mathbf{A}|\mathbf{x},I) \propto \int d\mathbf{s}\, P(\mathbf{x}|\mathbf{A},\mathbf{s},I)\, P(\mathbf{s}|I). \tag{5}$$

Second, the assumption that the mixing is noiseless, linear, and instantaneous, as described by Equation (1), is reflected in the assignment of a delta function for the likelihood, $P(\mathbf{x} | \mathbf{A}, \mathbf{s}, I)$. Third, the assumption that the source signals are statistically independent is reflected by factorizing the prior $P(\mathbf{s}|I)$, into the product of the priors of each independent source. Taking into account these assignments and using the Einstein summation convention to denote the matrix multiplication, we rewrite Equation (5) as

$$P(\mathbf{A}|\mathbf{x},I) \propto \int d\mathbf{s}\, \prod_i \delta(x_i - A_{ik} s_k) \prod_l p_l(s_l), \tag{6}$$

where $p_l(s_l)$ represents the prior probability of the amplitudes of source $l$. With a change of variables, $w_i = x_i - A_{ik} s_k$, the delta function allows us to evaluate the multi-dimensional integral,

$$P(\mathbf{A}|\mathbf{x},I) \propto \frac{1}{\det \mathbf{A}} \prod_l p_l(A^{-1}_{lk} x_k). \tag{7}$$

We have derived a formula, up to a normalization factor, for the probability that a given matrix $\mathbf{A}$, is the correct mixing matrix for our problem. It is now a matter of using the above formula to search for the most probable mixing matrix $\mathbf{A}$. With this in mind, we look at the logarithm of the probability

$$\log P(\mathbf{A}|\mathbf{x},I) = -\log \det \mathbf{A} + \sum_l \log p_l(A^{-1}_{lk} x_k) + C, \tag{8}$$

where $C$ is the logarithm of the normalization factor implicit in Equation (7).

To perform the separation, we are interested in the inverse of the mixing matrix, the separation matrix $\mathbf{W}$. Instead of searching for the mixing matrix $\mathbf{A}$, we search for the separation matrix $\mathbf{W}$, which maximizes the probability that $\mathbf{A}$ is the correct mixing matrix. A discussion of the arbitrariness of $\mathbf{A}$ and $\mathbf{W}$ in the blind source separation case can be found in a previous paper.[9] Rewriting Equation (8) in terms of $\mathbf{W}$, we get

$$\log P(\mathbf{A}|\mathbf{x},I) = \log \det \mathbf{W} + \sum_l \log p_l(W_{lk} x_k) + C. \tag{9}$$

To find the maximum of the logarithm of the posterior probability, $P(\mathbf{A} | \mathbf{x}, I)$, with respect to variation in the matrix $\mathbf{W}$, we take the derivative of Equation (9) with respect to the matrix elements $W_{ij}$

$$\frac{\partial}{\partial W_{ij}} \log P(\mathbf{A}|\mathbf{x},I) = \frac{\partial}{\partial W_{ij}} \left[ \log \det \mathbf{W} + \sum_l \log p_l(u_l) + C \right] \tag{10}$$

$$= A_{ji} + \frac{\partial}{\partial W_{ij}} \left[ \sum_l \log p_l(W_{lk} x_k) \right] \tag{11}$$

$$= A_{ji} + \frac{\partial u_m}{\partial W_{ij}} \frac{\partial}{\partial u_m} \left[ \sum_l \log p_l(u_l) \right] \tag{12}$$

$$= A_{ji} + \frac{\partial u_m}{\partial W_{ij}} \frac{\partial \log p_m(u_m)}{\partial u_m} \tag{13}$$

$$= A_{ji} + x_j \, \delta_{im} \, \frac{\partial \log p_m(u_m)}{\partial u_m} \tag{14}$$

$$\frac{\partial}{\partial W_{ij}} \log P(\mathbf{A} | \mathbf{x}, I) = A_{ji} + x_j \left( \frac{p'_i(u_i)}{p_i(u_i)} \right)_i \tag{15}$$

where we have introduced, $u_i = W_{ij} x_k$, and recognize that the ratio of the probability density of $u_i$ and its derivative is a column vector. The implementation of the Bell-Sejnowski ICA algorithm consists of writing Equation (15) in matrix form and using the data to perform a stochastic gradient ascent to find the separation matrix that maximizes Equation (9).

## 4. INCORPORATING SOURCE GEOMETRY AND SIGNAL PROPAGATION INFORMATION

As noted in the previous section, blind source separation deals with problems where there is a lack of prior information about the mixing process. In many physical applications this is not the case. We now demonstrate how the Bayesian formalism provides a natural and consistent means in which to incorporate prior knowledge about the mixing process.

Prior knowledge about a particular mixing process is expressed by the prior probability term $P(\mathbf{A} | I)$, in Equation (4). For the purpose of demonstration, we will choose a simple non-physical mixing process that has aspects of both the acoustic mixing problem and the neuroelectromagnetic mixing problem. We will assume that the sources radiate spherically with the intensity dropping off with the inverse square of the distance between the detector and the source. In addition, we will assume that there is no time delay for the signal to reach the detector, so that the mixing is instantaneous. Note that the radiation pattern is a good first approximation for sound sources, but that the lack of a delay is non-physical. While the lack of delay is appropriate for neuroelectromagnetic sources, the radiation pattern is not appropriate as neural sources produce potentials and fields that drop off as the inverse cube of the distance between the source and detector and exhibit directional dependence.

Given the assumptions regarding the mixing process, we expect that an element of the mixing matrix can be written in the form

$$A_{ij} = \frac{a_j}{4\pi |r_{ij}|^2} \tag{16}$$

where $|r_{ij}|$ is the distance between source $j$ and detector $i$, and $a_j$ is the amplitude of the source. The amplitude is included here because the source model that is typically used, Equation (38), does not explicitly include the amplitude of the source as a hyperparameter.

We can express the prior probability of the matrix $\mathbf{A}$ in terms of the prior probabilities of its elements. The easiest way to accomplish this is to assume that the elements of $\mathbf{A}$ are independent. This is not completely accurate, however, since knowledge about the distance between source $j$ and detector $i$ provides some information about the distance between source $j$ and detector $k$ when the detector positions relative to one another are known. This prior probability can be expressed as

$$P(\mathbf{A} | I) \approx \prod_{i,j} P(A_{ij} | I), \tag{17}$$

where $P(\mathbf{A} | I)$ is the prior probability of the mixing matrix and $P(A_{ij} | I)$ is the prior probability of a matrix element.

It is essential at this point to examine our model of the separation problem and to decide on which aspects to focus our attention. The model consists of a set of sources with source behavior, $\{s_j(t)\}$, compactly written as a vector of time series, $\mathbf{s}(t)$, with their amplitudes explicitly described by the set $\{a_j\}$. In addition, the sources have positions in space denoted by the set $\{\bar{s}_j\}$, and their distances relative to the detector positions $\{\bar{d}_i\}$, are given by $|r_{ij}|$. Using Equation (16), these parameters can be combined to form the mixing matrix $\mathbf{A}$, which describes the linear mixing in Equation (1). We may instead choose to focus our attention on this matrix $\mathbf{A}$, or its inverse.

As described in the previous section, to develop an algorithm, one needs to write the probability of the model parameters of interest conditional on the data and the prior information. If mixing of the signals is of primary interest, one may opt to work with $P(\mathbf{A} \mid \mathbf{x}, I)$. Whereas, if localization of the sources is of primary interest, one can work with $P(\mathbf{s} \mid \mathbf{x}, I)$, or $P(\mathbf{s}, \mathbf{a} \mid \mathbf{x}, I)$. In the latter case, one ends up performing a search where the parameters affect the solution both linearly and nonlinearly. It is possible to work with the linear problem, $P(\mathbf{A} \mid \mathbf{x}, I)$, and then use the results along with the prior information to estimate the source positions and amplitudes from the most probable matrix $\mathbf{A}$. We will follow this course.

Since our prior knowledge usually consists of information regarding the source geometry and perhaps source amplitude, we need to relate prior probabilities of the elements of the mixing matrix to prior probabilities of source location and amplitude. This is most easily performed by rewriting the joint probability of the model as

$$P(A_{ij}, a_j, |r_{ij}| \mid I) = P(A_{ij} \mid a_j, |r_{ij}|, I) P(a_j \mid I) P(|r_{ij}| \mid I) \tag{18}$$

and by marginalizing over the source amplitude and the distance between the source and detector

$$P(A_{ij} \mid I) = \int_0^\infty da_j \, P(a_j \mid I) \int_0^\infty d|r_{ij}| \, P(A_{ij} \mid a_j, |r_{ij}|, I) P(|r_{ij}| \mid I). \tag{19}$$

The integrals go from zero to infinity since both the source amplitudes and the distances from the sources to the detectors are positive. The term, $P(A_{ij} \mid a_j, |r_{ij}|, I)$ can be assigned a delta function derived from our belief that Equation (16) very accurately describes the signal propagation and that noise is negligible:

$$P(A_{ij} \mid I) = \int_0^\infty da_j \, P(a_j \mid I) \int_0^\infty d|r_{ij}| \, \delta\left( A_{ij} - \frac{a_j}{4\pi |r_{ij}|^2} \right) P(|r_{ij}| \mid I). \tag{20}$$

The integral over $|r_{ij}|$ can be evaluated, resulting in

$$P(A_{ij} \mid I) = \sqrt{\pi} \, A_{ij}^{-\frac{3}{2}} \int_0^\infty da_j \, \sqrt{a_j} \, P(a_j \mid I) P(|r_{ij}| \mid I) \tag{21}$$

where the first prior probability in the integral is the prior probability of the source amplitude, and the second is the prior probability of the source distance from the detector, where $|r_{ij}| = \sqrt{\dfrac{4\pi A_{ij}}{a_j}}$ .

In our treatment, we will assign a uniform distribution to the source amplitude prior:

$$P(a_j \mid I) = \begin{array}{ll} (b_2 - b_1)^{-1} & \text{for } b_1 \leq a_j \leq b_2 \\ 0 & \text{otherwise} \end{array} \tag{22}$$

where $b_1$ and $b_2$ are the smallest and largest possible values of $a_j$. To quantify our knowledge about the distances between the sources and detectors, we need to take into account our knowledge about the relative positions of the sources and detectors. If we have some information on the mean and the variance of the source positions, then the principle of maximum entropy dictates that we should use a Gaussian prior to describe this knowledge. However, we need a prior probability representing our prior information regarding the distance between a source and a detector, not the source position. Following a similar procedure as above in Equations (18) - (21), we obtain the following MaxEnt prior:

$$P(|r_{ij}| \mid I) = \frac{2\sqrt{2\pi} |r_{ij}| \sigma_j}{|\vec{d}_i - \vec{s}_j|} Exp\left[ \frac{-(|r_{ij}|^2 + |\vec{d}_i - \vec{s}_j|^2)}{2\sigma_j^2} \right] Sinh\left[ \frac{|r_{ij}| \cdot |\vec{d}_i - \vec{s}_j|}{\sigma_j^2} \right], \tag{23}$$

where $\vec{d}_i$ represents the position of detector $i$, $\vec{s}_j$ represents the mean believed position of source $j$, and $s_j^2$ represents the variance of the mean position. However, to simplify the final algorithm, we instead choose a similar prior that will allow us more easily integrate Equation (21). We choose the Gamma prior,

$$P(|r_{ij}||I) = \frac{\beta^{-\alpha} e^{\frac{|r_{ij}|}{\beta}} |r_{ij}|^{-1+\alpha}}{\Gamma(\alpha)}, \quad (24)$$

where the mean of the distance between detector $i$ and source $j$ is given by $|\vec{d}_i - \vec{s}_j| = \alpha\beta$, and the variance by $s_j^2 = \alpha\beta^2$. Aside from the advantage of allowing us to obtain an analytical solution of (21), the gamma prior also disallows negative values of distance and has a form similar to the MaxEnt prior in Equation (23).

Substituting these priors, (22) and (24), into the integral in (21), we get

$$P(A_{ij} | I) = \sqrt{\pi} A_{ij}^{-\frac{3}{2}} \int_0^\infty da_j \sqrt{a_j} (b_{2j} - b_{1j})^{-1} \frac{\beta_{ij}^{-\alpha_{ij}}}{\Gamma(\alpha_{ij})} e^{-\frac{1}{\beta_{ij}}\left(\frac{4\pi A_{ij}}{a_j}\right)^{-\frac{1}{2}}} \left(\frac{4\pi A_{ij}}{a_j}\right)^{\frac{-1+\alpha_{ij}}{2}}, \quad (25)$$

where it is understood that $b_{1j}$ and $b_{2j}$ refer to source $j$, and that $\alpha_{ij}$ and $\beta_{ij}$ are derived from the mean distance between source $j$ and detector $i$ and the variance of the source positions. Carrying out the final integration above, we obtain

$$P(A_{ij} | I) = \frac{4\pi}{\Gamma(\alpha_{ij})\beta_{ij}^2 (b_{2j} - b_{1j})} \left[ \gamma\left(2 + \alpha_{ij}, \frac{1}{\beta_{ij}}\sqrt{\frac{b_{1j}}{4\pi A_{ij}}}\right) - \gamma\left(2 + \alpha_{ij}, \frac{1}{\beta_{ij}}\sqrt{\frac{b_{2j}}{4\pi A_{ij}}}\right) \right] \quad (26)$$

where $\gamma(\bullet, \bullet)$ is the incomplete gamma function.[10] This in conjunction with Equation (17) provides a prior probability density for the mixing matrix **A**, which takes into account our knowledge of the amplitude of the sources, the geometry of the sources and the detectors, and the nature of the signal propagation.

This derived prior for the mixing matrix can be used in Equation (4). Following the derivation of the Bell-Sejnowski ICA algorithm, we compute the derivative of the logarithm of the probability, $P(\mathbf{A} | \mathbf{x}, I)$:

$$\frac{\partial}{\partial W_{ij}} \log P(\mathbf{A} | \mathbf{x}, I) = \frac{\partial}{\partial W_{ij}} \left[ \log \det \mathbf{W} + \sum_l \log p_l(W_{lk} x_k) + \sum_{k,l} \log P(A_{kl} | I) - C \right] \quad (27)$$

$$= A_{ji} + x_j \left(\frac{p_i'(u_i)}{p_i(u_i)}\right)_i + \frac{\partial A_{mn}}{\partial W_{ij}} \frac{\partial}{\partial A_{mn}} \left[ \sum_{k,l} \log P(A_{kl} | I) \right]. \quad (28)$$

Concentrating on the derivative of the last term

$$\frac{\partial}{\partial A_{mn}} \left[ \sum_{k,l} \log P(A_{kl} | I) \right] = \frac{\frac{\partial}{\partial A_{mn}} P(A_{mn} | I)}{P(A_{mn} | I)} \quad (29)$$

$$= \frac{\frac{\partial}{\partial A_{mn}} \left[ \gamma\left(2 + \alpha_{mn}, \frac{1}{\beta_{mn}}\sqrt{\frac{b_{1n}}{4\pi A_{mn}}}\right) - \gamma\left(2 + \alpha_{mn}, \frac{1}{\beta_{mn}}\sqrt{\frac{b_{2n}}{4\pi A_{mn}}}\right) \right]}{\left[ \gamma\left(2 + \alpha_{mn}, \frac{1}{\beta_{mn}}\sqrt{\frac{b_{1n}}{4\pi A_{mn}}}\right) - \gamma\left(2 + \alpha_{mn}, \frac{1}{\beta_{mn}}\sqrt{\frac{b_{2n}}{4\pi A_{mn}}}\right) \right]} \quad (30)$$

$$\frac{\partial}{\partial A_{mn}}\left[\sum_{k,l}\log P(A_{kl}\,|\,I)\right] = \frac{\left(\frac{1}{\beta_{mn}}\sqrt{\frac{b_{1n}}{4\pi A_{mn}}}\right)^{2+\alpha_{mn}} e^{-\frac{1}{\beta_{mn}}\sqrt{\frac{b_{1n}}{4\pi A_{mn}}}} - \left(\frac{1}{\beta_{mn}}\sqrt{\frac{b_{2n}}{4\pi A_{mn}}}\right)^{2+\alpha_{mn}} e^{-\frac{1}{\beta_{mn}}\sqrt{\frac{b_{2n}}{4\pi A_{mn}}}}}{2 A_{mn}\left[\gamma\left(2+\alpha_{mn},\frac{1}{\beta_{mn}}\sqrt{\frac{b_{1n}}{4\pi A_{mn}}}\right)-\gamma\left(2+\alpha_{mn},\frac{1}{\beta_{mn}}\sqrt{\frac{b_{2n}}{4\pi A_{mn}}}\right)\right]}. \quad (31)$$

Applying the above formula to each of the matrix elements $A_{mn}$, we obtain a matrix that we shall denote by **M**. Finally, we use the fact that the derivative of the mixing matrix **A**, with respect to its inverse, the separation matrix **W**, is given by

$$\frac{\partial A_{mn}}{\partial W_{ij}} = \frac{\partial A_{mn}}{\partial A_{ij}^{-1}} = -A_{mi}\,A_{jn} \quad (32)$$

to express Equation (28) as

$$\frac{\partial}{\partial W_{ij}}\log P(\mathbf{A}\,|\,x,I) = A_{ji} + x_j\left(\frac{p'_i(u_i)}{p_i(u_i)}\right)_i - A_{mi}\,A_{jn}\,M_{mn}, \quad (33)$$

where we are using the Einstein summation convention and the matrix element $M_{mn}$ is given by Equation (31). Equation (33) can be written in matrix form as

$$\frac{\partial}{\partial \mathbf{W}}\log P(\mathbf{A}\,|\,x,I) = \mathbf{A}^T + \left(\frac{p'_i(u_i)}{p_i(u_i)}\right)\mathbf{x}^T - \mathbf{A}^T\mathbf{M}\mathbf{A}^T. \quad (34)$$

As with the Bell-Sejnowski algorithm, one can easily implement a stochastic gradient search algorithm to find the separation matrix **W**, which satisfies the maximum a posteriori criterion. An initial guess for the separation matrix can be updated by setting $\Delta\mathbf{W}$ equal to Equation (34) above. It should be noted, however, that the equation is not covariant, since the object is not a matrix, but is the derivative of a scalar with respect to a matrix. The equation can be made covariant by transforming the gradient using an appropriate metric. It has been shown,[8, 11] that this can be accomplished by post-multiplying by $\mathbf{W}^T\mathbf{W}$. This results in the final form of the stochastic gradient update term,

$$\Delta\mathbf{W} = \mathbf{W} + \left(\frac{p'_i(u_i)}{p_i(u_i)}\right)\mathbf{u}^T\mathbf{W} - \mathbf{A}^T\mathbf{M}\mathbf{W} \quad (35)$$

The update rule for the stochastic gradient algorithm is implemented using

$$\mathbf{W}_{i+1} = \mathbf{W}_i + l\,\Delta\mathbf{W} \quad (36)$$

where $l$ is the learning rate, and $\Delta\mathbf{W}$ is obtained by applying Equation (35) to a sample of the data.

## 5. DEMONSTRATION OF THE SEPARATION ALGORITHM

The separation algorithm is applicable when the researcher has some prior information or belief regarding the positions of the sources and their amplitudes. Often one has information in the form of a mean source position $\vec{s}_j$, for a source $j$ and an associated variance of that mean $\mathsf{s}_j^2$. The mean prior source position is used to determine the mean distance between the $i^{th}$ detector and the $j^{th}$ source, $\mu_{ij} = |\vec{d}_i - \vec{s}_j|$. The parameters of the Gamma prior representing our knowledge about the distances between the detectors and the sources in Equation (24) are found by $\alpha_{ij} = \frac{\mu_{ij}^2}{\sigma_j^2}$ and $\beta_{ij} = \frac{\sigma_j^2}{\mu_{ij}}$. The cutoffs for the amplitudes of each of the sources are chosen with $b_{1j}$ as the minimum possible amplitude of source $j$ and $b_{2j}$ as the maximum possible amplitude of source $j$, measured in units of the characteristic width of the prior source amplitude density.

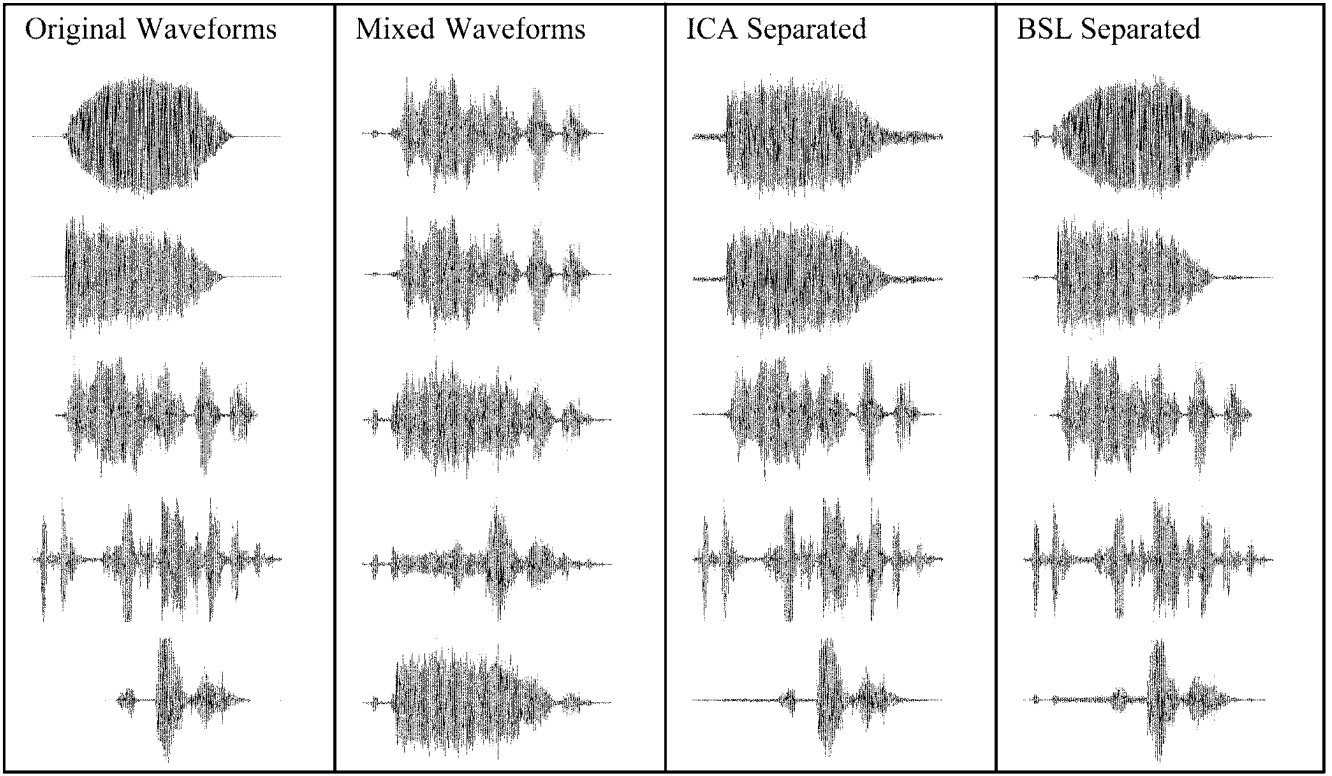

**Figure 1.** Displayed are the original source waveforms, the artificially mixed waveforms, the results from ICA, and the results from BSL. From top to bottom, the original waveforms are Hail (a two-toned whistle), Photon Torpedo (a noisy frequency glide), and the spoken phrases "Live long and prosper", "Captain we're losing power in the warp engines", and "He's dead Jim". The mixtures are not as uniform as in randomly mixed signals due to the inverse square law of propagation. While ICA accurately separates the speech sounds, it cannot separate the Hail and Photon Torpedo. The result is a diagonal solution where the top waveform is a sum of the two original signals and the second is the difference. The additional information used by BSL allows separation of all of the sounds, however the inaccurate source amplitude density for the Hail and Photon Torpedo causes a small amount of mixing of the speech sounds.

The final prior to assign is that of the prior density of the source amplitudes. Speech sounds typically exhibit amplitude histograms that are unimodal and hyper-Gaussian. It has been shown that the sigmoid function is an effective source amplitude prior distribution for speech sounds,[2]

$$g(u) = \frac{1}{1+e^{-u}}, \qquad (37)$$

which has a hyper-Gaussian probability density given by

$$p_i(u_i) = \frac{dg(u_i)}{du_i} = g(u_i)(1 - g(u_i)). \qquad (38)$$

This results in the column vector in Equation (35) having elements given by

$$\left(\frac{p'_i(u_i)}{p_i(u_i)}\right) = 1 - 2g(u_i). \qquad (39)$$

With all of the priors explicitly defined, the stochastic gradient ascent can be performed to find the separation matrix **W**, by applying the algorithm described by Equations (35), (36) and (39).

We now demonstrate the BSL algorithm on a set of artificially mixed signals. The five source signals were derived from the TV show Star Trek. Three of the source waveforms are speech and the other two are a Hail, which is similar to a two-tone whistle, and a Photon Torpedo Blast, which is a noisy frequency glide. The detectors and sources were randomly placed in a cubic volume with side length 10. The source signals were artificially mixed using a mixing matrix derived from Equation (16). The variances of the prior source positions were chosen randomly from a uniform distribution ranging from 0 to 1. The prior source positions were randomly chosen from a normal distribution using the correct position as the mean and previously determined variance. The cutoffs for the uniform amplitude distributions were the estimated amplitudes in units of the characteristic width of the sigmoid function ± 4.0.

In Figure 1 we compare the results obtained for the separation of the Hail and Photon Torpedo sounds using both ICA and BSL. As previously described,[9] ICA cannot accurately separate the Hail and Photon Torpedo due to the fact that the source amplitude prior is very sharply peaked, whereas the Hail and Photon Torpedo amplitude histograms are very broad. This lack of accurate information results in a diagonal solution in which the two separated signals consist of the sum and difference of re-scaled versions of the original source waveforms. In contrast, BSL is able to use information regarding the source locations and the way in which the signals propagate to the detectors to offset the inaccurate source amplitude information. This inaccurate source amplitude density information, however, still causes problems by slightly mixing the speech sounds.

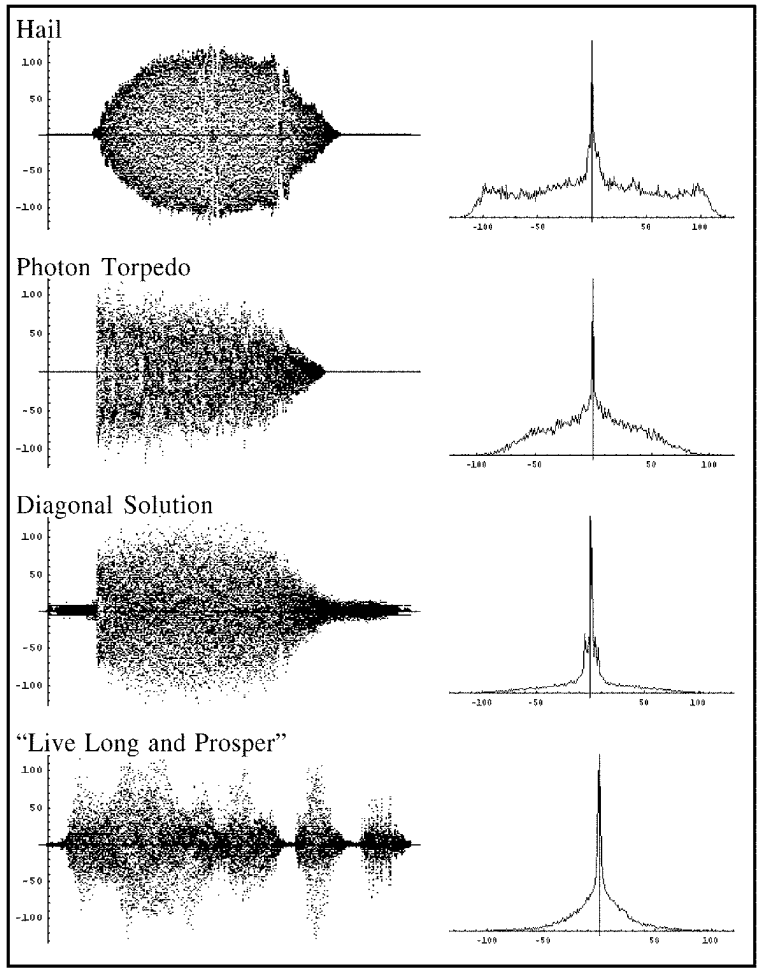

**Figure 2.** Displayed are four sample waveforms and their amplitude histograms. Note that the Hail and Photon Torpedo has source amplitude densities that are very broad, whereas the Diagonal Solution and the speech are hyper-Gaussian.

In Figure 2 we show the amplitude histograms of the Hail, Photon Torpedo, one of the ICA diagonal solutions and the spoken phrase "Live Long and Prosper". Note that the amplitude histograms of the speech sound and the diagonal solution are both very sharply peaked, or hyper-Gaussian. The Hail and Photon Torpedo, on the other hand, have much broader amplitude densities. By mixing the Hail and Photon Torpedo sounds to form a diagonal solution, a hyper-Gaussian density is attained that more closely matches the sigmoid prior.

## 6. ESTIMATION OF SOURCE PARAMETERS

In this section we will briefly demonstrate how source parameters, such as the source position, can be estimated from the optimal mixing matrix. The basic idea is again to use Bayes' Theorem and marginalization to write the probability of the estimated source positions $\hat{s}_j$ as

$$P(\hat{s}_j | A_j, I) = \int da_j \, P(A_j | \hat{s}_j, a_j, I) \, P(\hat{s}_j | I) \, P(a_j | I) , \qquad (40)$$

where $A_j$ represents the $j^{th}$ column of the mixing matrix **A**. Since we assumed earlier that the matrix elements $A_{ij}$, are independent, we can express the likelihood term in Equation (40) as a product of the likelihoods for each element in the $j^{th}$ column of **A**:

$$P(A_j | \hat{s}_j, a_j, I) = \prod_{i=1}^{n} P(A_{ij} | \hat{s}_j, a_j, I). \tag{41}$$

If we have some information regarding the precision of our estimate of the mixing matrix **A**, we can assign a Gaussian density to represent the likelihoods in Equation (41). The variance of **A**, which represents this precision, can be estimated by looking at the second derivative of the logarithm of the posterior probability of **A** (Equation (27)). The assignment of a Gaussian prior to the source position prior is appropriate when the source's mean position and variance are known. Finally, Equation (22) describes the uniform prior we assigned to the source amplitudes. Using these probabilities in Equation (40) we can evaluate the probability that any position $\hat{s}_j$, is the correct position for source $j$:

$$P(\hat{s}_j | A_j, I) = P(\hat{s}_j | I) \int_{b_{1j}}^{b_{2j}} da_j \prod_{i=1}^{n} P(A_{ij} | \hat{s}_j, a_j, I)(b_{2j} - b_{1j})^{-1} \tag{42}$$

$$P(\hat{s}_j | A_j, I) = \frac{1}{\sqrt{2\pi}\,\sigma_j} Exp\left(-\frac{|\hat{s}_j - \bar{s}_j|^2}{2\sigma_j^2}\right)(b_{2j} - b_{1j})^{-1} \int_{b_{1j}}^{b_{2j}} da_j \prod_{i=1}^{n} \frac{1}{\sqrt{2\pi}\,\sigma_{ij}} Exp\left(-\frac{1}{2\sigma_{ij}^2}\left(A_{ij} - \frac{a_j}{4\pi|\vec{d}_i - \hat{s}_j|^2}\right)^2\right), \tag{43}$$

where $s_j^2$ represents the variance of the mean prior source position, $\bar{s}_j$, and $s_{ij}^2$ represents the variance in our estimated value of the element $A_{ij}$, of the mixing matrix. To implement a search, it is more efficient to find the maximum of the logarithm of Equation (43).

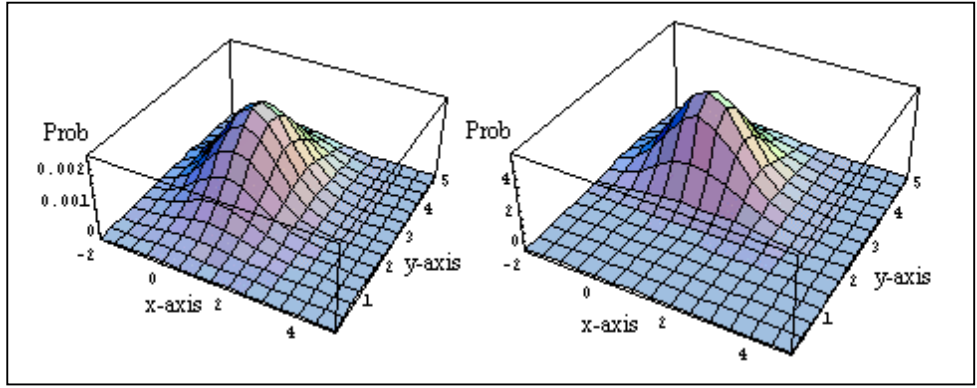

**Figure 3.** These graphs show the probability of the position for the Hail source calculated using Equation (43). In (a) the believed variance of the mixing matrix elements was set to 5.0, and in (b) the variance was set to 0.1. Notice that when we believe that our mixing matrix is imprecise, the posterior probability is dominated by the Gaussian prior probability. However, if we believe that the mixing matrix elements are very precise then we observe that the overall probability of the maximum of the distribution is increased significantly. In addition, locations near the detectors are excluded as being possible sites for the source.

Instead of actually locating the source positions in an experiment, we plot the posterior probability of the source position. This allows one to visualize the most likely position for the source and the uncertainty of our solution. We now consider an experiment that involved only the mixture of the Hail and the Photon Torpedo sounds. The source space is two-dimensional because with only two detectors there is an axial symmetry.  We thus constraint the source positions to lie in the xy-plane.  The detectors were located on the x-axis at (1,0) and (-1,0).  The Hail source is located at (3, 0.5) and the Photon Torpedo at (2, -0.75).  The variance in the believed source positions was chosen to be 1.0.  The uniform amplitude prior extended from 10 to 15 for the Photon Torpedo and 15 to 25 for the Hail.  The true amplitude of the Photon Torpedo is estimated to be about 13 and the Hail to be about 21.  The algorithm separated the signals with a percent error of 13.4% for the Hail and 36.4% for the Photon Torpedo.

In Figure 3, we plot the probability of the position of the Hail source.  For the purposes of illustration, we did not use the estimated variance of the values of the mixing matrix to calculate these probabilities.  Instead, Figure 3a shows the posterior probability of the position of the Hail source given that we believe our values of the separation matrix elements to be imprecise.  Note that this distribution is dominated by the Gaussian source position prior because the additional information regarding the mixing of the signals was deemed inaccurate.  In Figure 3b we repeat the calculation, but this time assuming the mixing matrix elements are very accurate.  This additional information describing the way the signals were mixed significantly modifies the posterior probability.  The probability density describing the possible position of the source is now more precisely localized and has a much greater magnitude.

## 7. SUMMARY

We have demonstrated that the Bayesian methodology provides a natural and logically consistent means by which prior information can be incorporated into a specific source separation problem.  An artificial mixing problem that exhibits features present in both the acoustic and neuroelectromagnetic separation problems was used to demonstrate the methodology.  Although, the specific algorithm derived in this paper is probably not particularly useful, our intention was to demonstrate how a specific separation algorithm can be constructed.

The algorithm was demonstrated on a set of sound signals artificially mixed by randomly choosing source and detector locations in space and by demanding that the signal amplitude falls off as the inverse square of the distance from the source.  Both the current BSL algorithm and ICA were used to separate the signals.  While both algorithms made the same assumptions regarding the source amplitude densities, the BSL algorithm incorporated additional information regarding the nature of the signal propagation, the amplitude of the signals and the believed positions of the sources.  While this additional information allowed BSL to separate the Hail and Photon Torpedo sounds, which are not separable under these conditions by ICA, the quality of the separated speech sounds suffered slightly.  The important point here is that the incorporation of additional information is not a substitute for the improvement of inaccurate information.  For the separation of sounds like the Hail and Photon Torpedo, one would do better to use less precise, but accurate, source amplitude density information and to include additional information regarding the source and detector positions and the nature of the signal propagation.

While the hyper-Gaussian source priors have been shown to be very useful in describing our knowledge about the amplitude densities of speech signals,[2,7] they have been shown to be inaccurate for describing slowly-varying waveforms such as those recorded using EEG and MEG.[9]  These slowly varying waveforms have source amplitude densities that are multi-modal in form and require several hyperparameters to model.  In the case of neuroelectrophysiologic signals, where these amplitude densities are difficult to model, it will be advantageous to incorporate as much additional information as possible.  In this case, additional information could include the possible source locations and orientations, the nature of the signal propagation, as well as any information about the nature of the signals themselves.[3]

It is often true in Bayesian statistics that very simple information can have a tremendous impact on the ability to infer a solution.  The success of ICA is an excellent example of this situation.  It is critical for those researchers working with specific source separation problems to evaluate the importance and effect of each piece of additional information.  Bayesian solutions have a tendency to become very difficult, mainly due to the need to marginalize over complicated priors.  The exclusion of information that is not found to be particularly helpful in a solution may be as important as inclusion of some overlooked details relevant to the problem.

We conclude by discussing some important points regarding the specific BSL algorithm derived in this paper.  First, by assuming independence of the elements of the mixing matrix, Equation (17), we ignore the symmetries imposed by the geometrical configuration of the detectors.  Inclusion of this information may dramatically improve the performance of the algorithm.  Second, if one has access to information about the mean positions of the sources and their respective variances, one should assign the MaxEnt prior, Equation (23), to represent the prior probability of the distances between the sources and detectors.  By assigning the Gamma prior instead of the MaxEnt prior, an unknown bias is introduced into the problem.  A comparison of the performance of these priors should be performed since the MaxEnt prior may require significantly more numerical computations per iteration step.  Finally, the BSL algorithm developed here is designed to find the most probable mixing matrix.  It should be noted that the most probable mixing matrix does not correspond to the most probable separation matrix, or the most probable source locations.  This is due to the way in which probability densities transform under a change of variables. An excellent example of this occurs in the context of blackbody radiation in physics. The wavelength of light in blackbody radiation with the greatest energy density does not correspond to the frequency of light with the greatest energy density.

## 8. ACKNOWLEDGEMENTS

The author would like to thank Irv Hochberg and Herb Vaughan for their advice and support. This work was supported in part by Albert Einstein College of Medicine, The City University of New York Graduate Center, and NIH NIDCD 5 T32 DC00039-05.

## 9. REFERENCES

1. E.C. Cherry, "Some experiments on the recognition of speech, with one and with two ears", *J. Acoust. Soc. Am.* **25**(5), pp. 975-979, 1953.
2. A.J. Bell and T.J. Sejnowski, "An information-maximization approach to blind separation and blind deconvolution", *Neural Comp.* **7**, pp. 1129-1159, 1995.
3. K.H. Knuth and H.G. Vaughan, Jr., "The Bayesian origin of blind source separation and electromagnetic source estimation", To be published in *Maximum Entropy and Bayesian Methods, Munich 1998*.
4. J.-F. Cardoso, "Infomax and maximum likelihood for source separation", *IEEE Lectures on Signal Processing* **4**(4), pp. 112-114, 1997.
5. P. Comon, "Independent component analysis, a new concept?", *Signal Processing* **36**, pp. 287-314, 1994.
6. H.H. Yang and S. Amari, "Adaptive on-line learning algorithms for blind separation - maximum entropy and minimum mutual information", *Neural Comp.* **9**, pp. 1457-1482, 1997.
7. B.A. Pearlmutter and L.C. Parra, "A context-sensitive generalization of ICA", *1996 International Conference on neural Information Processing, Hong Kong*, 1996.
8. D.J.C. MacKay, "Maximum likelihood and covariant algorithms for independent component analysis", Draft Paper, http://wol.ra.phy.cam.ac.uk/mackay/, 1996.
9. K.H. Knuth, "Difficulties applying recent blind source separation techniques to EEG and MEG", To be published in *Maximum Entropy and Bayesian Methods, Boise 1997*, 1998.
10. M. Abramowitz and I.A. Stegun, *Handbook of mathematical functions*, p. 260, Dover Publications Inc., NY, 1972.
11. S. Amari, "Natural gradient works efficiently in learning", *Neural Comp.* **10**, pp.251-276, 1998.